\documentclass[9pt,twocolumn,twoside,]{pinp}

\providecommand{\tightlist}{%
  \setlength{\itemsep}{0pt}\setlength{\parskip}{0pt}}

\usepackage[T1]{fontenc}
\usepackage[utf8]{inputenc}

\definecolor{pinpblue}{HTML}{185FAF}  
\definecolor{pnasbluetext}{RGB}{101,0,0} %

 \newcommand{\pkg}[1]{\textbf{#1}}

\title{Thirteen Simple Steps for Creating An R Package with an External C++
Library}

\author[1]{Dirk Eddelbuettel}

  \affil[1]{Department of Statistics, University of Illinois, Urbana-Champaign, IL,
USA}

\leadauthor{Eddelbuettel}

\begin{abstract}
We desribe how we extend R with an external C++ code library by using
the Rcpp package. Our working example uses the recent machine learning
library and application `Corels' providing optimal yet easily
interpretable rule lists \citep{arxiv:corels} which we bring to R in the
form of the \pkg{RcppCorels} package \citep{github:rcppcorels}. We
discuss each step in the process, and derive a set of simple rules and
recommendations which are illustrated with the concrete example.
\end{abstract}

\dates{This version was compiled on \today} 

\pinpfootercontents{Thirteen Steps for R and C++ Library Packages}

\begin{document}

\verticaladjustment{-2pt}

\maketitle
\thispagestyle{firststyle}
\ifthenelse{\boolean{shortarticle}}{\ifthenelse{\boolean{singlecolumn}}{\abscontentformatted}{\abscontent}}{}


\hypertarget{introduction}{%
\section{Introduction}\label{introduction}}

The process of building a new package with Rcpp can range from the very
easy---a single simple C++ function---to the very complex. If, and how,
external resources are utilised makes a big difference as this too can
range from the very simple---making use of a header-only library, or
directly including a few C++ source files without further
dependencies---to the very complex.

Yet a lot of the important action happens in the middle ground. Packages
may bring their own source code, but also depend on just one or two
external libraries. This paper describes one such approach in detail:
how we turned the Corels application \citep{arxiv:corels,github:corels}
(provided as a standalone C++-based executable) into an R-callable
package \textbf{RcppCorels} \citep{github:rcppcorels} via \textbf{Rcpp}
\citep{CRAN:Rcpp,JSS:Rcpp}.

\hypertarget{the-thirteen-key-steps}{%
\section{The Thirteen Key Steps}\label{the-thirteen-key-steps}}

\hypertarget{ensure-use-of-a-suitable-license}{%
\subsection{Ensure Use of a Suitable
license}\label{ensure-use-of-a-suitable-license}}

Before embarking on such a journey, it is best to ensure that the
licensing framework is suitable. Many different open-source licenses
exists, yet a few key ones dominate and can generally be used \emph{with
each other}. There is however a fair amount of possible legalese
involved, so it is useful to check inter-license compatibility, as well
as general usability of the license in question. Several sites can help
via license recommendations, and checks for interoperability. One
example is the site at
\href{https://choosealicense.com/}{choosealicense.com} (which is backed
by GitHub) can help, as can
\href{https://tldrlegal.com/}{tldrlegal.com}. License choice is a
complex topic, and general recommendations are difficult to make besides
the key point of sticking to already-established and known licenses.

\hypertarget{ensure-the-software-builds}{%
\subsection{Ensure the Software
builds}\label{ensure-the-software-builds}}

In order to see how hard it may to combine an external entity, either a
program a library, with R, it helps to ensure that the external entity
actually still builds and runs.

This may seem like a small and obvious steps, but experience suggests
that it worth asserting the ability to build with current tools, and
possibky also with more than one compiler or build-system. Consideration
to other platforms used by R also matter a great deal as one of the
strengths of the R package system is its ability to cover the three key
operating system families.

\hypertarget{ensure-it-still-works}{%
\subsection{Ensure it still works}\label{ensure-it-still-works}}

This may seem like a variation on the previous point, but besides the
ability to \emph{build} we also need to ensure the ability to \emph{run}
the software. If the external entity has tests and demo, it is highly
recommended to run them. If there are reference results, we should
ensure that they are still obtained, and also that the run-time
performance it still (at a minimum) reasonable.

\hypertarget{ensure-it-is-compelling}{%
\subsection{Ensure it is compelling}\label{ensure-it-is-compelling}}

This is of course a very basic litmus test: is the new software
relevant? Is is helpful? Would others benefit from having it packaged
and maintained?

\hypertarget{start-an-rcpp-package}{%
\subsection{Start an Rcpp package}\label{start-an-rcpp-package}}

The first step in getting a new package combing R and C++ is often the
creation of a new Rcpp package. There are several helper functions to
choose from. A natural first choice is \texttt{Rcpp.package.skeleton()}
from the \textbf{Rcpp} package \citep{CRAN:Rcpp}. It can be improved by
having the optional helper package \textbf{pkgKitten}
\citep{CRAN:pkgKitten} around as its \texttt{kitten()} function smoothes
some rougher edges left by the underlying Base R function
\texttt{package.skeleton()}. This step is shown below in then appendix,
and corresponds to the first commit, followed by a first edit of file
\texttt{DESCRIPTION}.

Any code added by the helper functions, often just a simple
\texttt{helloWorld()} variant, can be run to ensure that the package is
indeed functional. More importantly, at this stage, we can also start
building the package as a compressed tar archive and run the R checker
on it.

\hypertarget{integrate-external-package}{%
\subsection{Integrate External
Package}\label{integrate-external-package}}

Given a basic package with C++ support, we can now turn to integrating
the external package. This complexity of this step can, as alluded to
earlier, vary from very easy to very complex. Simple cases include just
dependending on library headers which can either be copied to the
package, or be provided by another package such as \textbf{BH}
\citep{CRAN:BH}. It may also be a dependency on a fairly standard
library available on most if not all systems. The graphics formats bmp,
jpeg or png may be example; text formats like JSON or XML are another.
One difficulty, though, may be that \emph{run-time} support does not
always guarantee \emph{compile-time} support. In these cases, a
\texttt{-dev} or \texttt{-devel} package may need to be installed.

In the concrete case of Corels, we

\begin{itemize}
\tightlist
\item
  copied all existing C++ source and header files over into the
  \texttt{src/} directory;
\item
  renamed all header files from \texttt{*.hh} to \texttt{*.h} to comply
  with an R preference;
\item
  create a minimal \texttt{src/Makevars} file, here with link
  instructions for GMP;
\item
  moved \texttt{main.cc} to a subdirectory as we cannot build with
  another \texttt{main()} function (and R will not include files from
  subdirectories);
\item
  added a minimal R-callable function along with a \texttt{logger}
  instance.
\end{itemize}

Here, the last step was needed as the file \texttt{main.cc} provided a
global instance referred to from other files. Hence, a minimal
R-callable wrapper is being added at this stage (shown in the appendix
as well). Actual functionality will be added later.

We will come back to the step concerning the link instructions.

As this point we have a package for R also containing the library we
want to add.

\hypertarget{make-the-external-code-compliant-with-r-policies}{%
\subsection{Make the External Code compliant with R
Policies}\label{make-the-external-code-compliant-with-r-policies}}

R has fairly strict guidelines, defined both in the \emph{CRAN
Repository Policy} document at the CRAN website, and in the manual
\emph{Writing R Extension}. Certain standard C and C++ functions are not
permitted as their use could interfere with running code from R. This
includes somewhat obvious recommendations (``do not call
\texttt{abort}'' as it would terminate the R sessions) but extends to
not using native print methods in order to cooperate better with the
input and output facilities of R. So here, and reflecting that last
aspect, we changed all calls to \texttt{printf()} to calls to
\texttt{Rprintf()}. Similarly, R prefers its own (well-tested)
random-number generators so we replaced one (scaled) call to
\texttt{random()\ /\ RAND\_MAX} with the equivalent call to R's
\texttt{unif\_rand()}. We also avoided one use of \texttt{stdout} in
\texttt{rulelib.h}.

The requirement for such changes may seem excessive at first, but the
value added stemming from consistent application of the CRAN Policies is
appreciated by most R users.

\hypertarget{complete-the-interface}{%
\subsection{Complete the Interface}\label{complete-the-interface}}

In order to further test the package, and of course also for actual use,
we need to expose the key parameters and arguments. Corels parsed
command-line arguments; we can translate this directly into suitable
arguments for the main function. At a first pass, we created the
following interface:

\begin{Shaded}
\begin{Highlighting}[]
\CommentTok{// [[Rcpp::export]]}
\DataTypeTok{bool}\NormalTok{ corels(}\BuiltInTok{std::}\NormalTok{string rules_file,}
            \BuiltInTok{std::}\NormalTok{string labels_file,}
            \BuiltInTok{std::}\NormalTok{string log_dir,}
            \BuiltInTok{std::}\NormalTok{string meta_file = }\StringTok{""}\NormalTok{,}
            \DataTypeTok{bool}\NormalTok{ run_bfs = }\KeywordTok{false}\NormalTok{,}
            \DataTypeTok{bool}\NormalTok{ calculate_size = }\KeywordTok{false}\NormalTok{,}
            \DataTypeTok{bool}\NormalTok{ run_curiosity = }\KeywordTok{false}\NormalTok{,}
            \DataTypeTok{int}\NormalTok{ curiosity_policy = }\DecValTok{0}\NormalTok{,}
            \DataTypeTok{bool}\NormalTok{ latex_out = }\KeywordTok{false}\NormalTok{,}
            \DataTypeTok{int} \DataTypeTok{map_type}\NormalTok{ = }\DecValTok{0}\NormalTok{,}
            \DataTypeTok{int}\NormalTok{ verbosity = }\DecValTok{0}\NormalTok{,}
            \DataTypeTok{int}\NormalTok{ max_num_nodes = }\DecValTok{100000}\NormalTok{,}
            \DataTypeTok{double}\NormalTok{ regularization = }\FloatTok{0.01}\NormalTok{,}
            \DataTypeTok{int}\NormalTok{ logging_frequency = }\DecValTok{1000}\NormalTok{,}
            \DataTypeTok{int}\NormalTok{ ablation = }\DecValTok{0}\NormalTok{) \{}

  \CommentTok{// actual function body omitted}
\NormalTok{\}}
\end{Highlighting}
\end{Shaded}

Rcpp facilities the integration by adding another wrapper exposing all
the function arguments, and setting up required arguments without
default (the first three) along with optional arguments given a default.
The user can now call \texttt{corels()} from R with three required
arguments (the two input files plus the log directory) as well as number
of optional arguments.

\hypertarget{add-sample-data}{%
\subsection{Add Sample Data}\label{add-sample-data}}

R package can access data files that are shipped with them. That is very
useful feature, and we therefore also copy in the files include in the
Corels repository and its \texttt{data/} directory.

\begin{Shaded}
\begin{Highlighting}[]
\NormalTok{fs}\OperatorTok{::}\KeywordTok{dir_tree}\NormalTok{(}\StringTok{"../rcppcorels/inst/sample_data"}\NormalTok{)}
\CommentTok{#  ../rcppcorels/inst/sample_data}
\CommentTok{#  +-- compas_test-binary.csv}
\CommentTok{#  +-- compas_test.csv}
\CommentTok{#  +-- compas_test.label}
\CommentTok{#  +-- compas_test.out}
\CommentTok{#  +-- compas_train-binary.csv}
\CommentTok{#  +-- compas_train.csv}
\CommentTok{#  +-- compas_train.label}
\CommentTok{#  +-- compas_train.minor}
\CommentTok{#  \textbackslash{}-- compas_train.out}
\end{Highlighting}
\end{Shaded}

\hypertarget{set-up-working-example}{%
\subsection{Set up working example}\label{set-up-working-example}}

Combining the two preceding steps, we can now offer an illustrative
example. It is included in the helpd page for function \texttt{corels()}
and can be run from R via \texttt{example("corels")}.

\begin{Shaded}
\begin{Highlighting}[]
\KeywordTok{library}\NormalTok{(RcppCorels)}

\NormalTok{.sysfile <-}\StringTok{ }\ControlFlowTok{function}\NormalTok{(f)   }\CommentTok{# helper function}
  \KeywordTok{system.file}\NormalTok{(}\StringTok{"sample_data"}\NormalTok{,f,}\DataTypeTok{package=}\StringTok{"RcppCorels"}\NormalTok{)}

\NormalTok{rules_file <-}\StringTok{ }\KeywordTok{.sysfile}\NormalTok{(}\StringTok{"compas_train.out"}\NormalTok{)}
\NormalTok{label_file <-}\StringTok{ }\KeywordTok{.sysfile}\NormalTok{(}\StringTok{"compas_train.label"}\NormalTok{)}
\NormalTok{meta_file <-}\StringTok{ }\KeywordTok{.sysfile}\NormalTok{(}\StringTok{"compas_train.minor"}\NormalTok{)}
\NormalTok{logdir <-}\StringTok{ }\KeywordTok{tempdir}\NormalTok{()    }

\KeywordTok{stopifnot}\NormalTok{(}\KeywordTok{file.exists}\NormalTok{(rules_file),}
          \KeywordTok{file.exists}\NormalTok{(labels_file),}
          \KeywordTok{file.exists}\NormalTok{(meta_file),}
          \KeywordTok{dir.exists}\NormalTok{(logdir))}

\KeywordTok{corels}\NormalTok{(rules_file, labels_file, logdir, meta_file,}
       \DataTypeTok{verbosity =} \DecValTok{100}\NormalTok{,}
       \DataTypeTok{regularization =} \FloatTok{0.015}\NormalTok{,}
       \DataTypeTok{curiosity_policy =} \DecValTok{2}\NormalTok{,   }\CommentTok{# by lower bound}
       \DataTypeTok{map_type =} \DecValTok{1}\NormalTok{)           }\CommentTok{# permutation map}

\KeywordTok{cat}\NormalTok{(}\StringTok{"See "}\NormalTok{, logdir, }\StringTok{" for result file."}\NormalTok{)}
\end{Highlighting}
\end{Shaded}

In the example, we pass the two required arguments for rules and labels
files, the optional argument for the `meta' file as well as an added
required argument for the output directory. R policy prohibits writing
in user-directories, we default to using the temporary directory of the
current session, and report its value at the end. For other arguments
default values are used.

\hypertarget{finesse-library-dependencies}{%
\subsection{Finesse Library
Dependencies}\label{finesse-library-dependencies}}

One fairly common difficulty in bringing a library to R via a package
consists of external dependencies. In the case of `Corels', the GNU GMP
library for multi-precision arithmetic is needed with both its C and C++
language bindings. In order to detect presence of a required (or maybe
optional library), tools like `autoconf' or `cmake' are often used. For
example, to detect presence of the the GNU GMP, CRAN package
\textbf{sbrl} \citep{CRAN:sbrl} uses `autoconf' and turns optional use
on if the library is present. Here, however, we need to test for both
the C and C++ library bindings.

One additional problem with `Corels' is that at present, compilation
depends on GMP. So while we can use `autoconf' to detect it, we have to
abort the build if the library (or its C++ parts) are not present.

\hypertarget{finalise-license-and-copyright}{%
\subsection{Finalise License and
Copyright}\label{finalise-license-and-copyright}}

It is good (and common) practice to clearly attribute authorship. Here,
credit is given to the `Corels' team and authors as well as to the
authors of the underlying `rulelib' code used by `Corels' via the file
\texttt{inst/AUTHORS} (which will be installed as \texttt{AUTHORS} with
the package. In addition, the file \texttt{inst/LICENSE} clarifies the
GNU GPL-3 license for `RcppCorels' and `Corels', and the MIT license for
`rulelib'.

\hypertarget{additional-bonus-some-more-meta-files}{%
\subsection{Additional Bonus: Some more `meta'
files}\label{additional-bonus-some-more-meta-files}}

Several files help to improve the package. For example,
\texttt{.Rbuildignore} allows to exclude listed files from the resulting
R package keeping it well-defined. Similarly, \texttt{.gitignore} can
exclude files from being added to the \texttt{git} repository. We also
like \texttt{.editorconfig} for consistent editing default across a
range of modern editors.

\hypertarget{summary}{%
\section{Summary}\label{summary}}

We describe s series of steps to turn the standalone library `Corels'
describes by \citet{arxiv:corels} into a R package \textbf{RcppCorels}
using the facilities offered by \textbf{Rcpp} \citep{CRAN:Rcpp}. Along
the way, we illustrate key aspects of the R package standards and CRAN
Repository Policy proving a template for other research software wishing
to provide their implementations in a form that is accessibly by R
users.

\bibliography{references}
\bibliographystyle{jss}

\newpage
\onecolumn

\hypertarget{appendix-1-creating-the-basic-package}{%
\subsection{Appendix 1: Creating the basic
package}\label{appendix-1-creating-the-basic-package}}

\begin{Shaded}
\begin{Highlighting}[]
\ExtensionTok{edd@rob}\NormalTok{:~/git$ r --packages Rcpp --eval }\StringTok{'Rcpp.package.skeleton("RcppCorels")'}

\ExtensionTok{Attaching}\NormalTok{ package: ‘utils’}

\ExtensionTok{The}\NormalTok{ following objects are masked from ‘package:Rcpp’:}

    \ExtensionTok{.DollarNames}\NormalTok{, prompt}

\ExtensionTok{Creating}\NormalTok{ directories ...}
\ExtensionTok{Creating}\NormalTok{ DESCRIPTION ...}
\ExtensionTok{Creating}\NormalTok{ NAMESPACE ...}
\ExtensionTok{Creating}\NormalTok{ Read-and-delete-me ...}
\ExtensionTok{Saving}\NormalTok{ functions and data ...}
\ExtensionTok{Making}\NormalTok{ help files ...}
\ExtensionTok{Done.}
\ExtensionTok{Further}\NormalTok{ steps are described in }\StringTok{'./RcppCorels/Read-and-delete-me'}\NormalTok{.}

\ExtensionTok{Adding}\NormalTok{ Rcpp settings}
 \OperatorTok{>>} \ExtensionTok{added}\NormalTok{ Imports: Rcpp}
 \OperatorTok{>>} \ExtensionTok{added}\NormalTok{ LinkingTo: Rcpp}
 \OperatorTok{>>} \ExtensionTok{added}\NormalTok{ useDynLib directive to NAMESPACE}
 \OperatorTok{>>} \ExtensionTok{added}\NormalTok{ importFrom(Rcpp, evalCpp) }\ExtensionTok{directive}\NormalTok{ to NAMESPACE}
 \OperatorTok{>>} \ExtensionTok{added}\NormalTok{ example src file using Rcpp attributes}
 \OperatorTok{>>} \ExtensionTok{added}\NormalTok{ Rd file for rcpp_hello_world}
 \OperatorTok{>>} \ExtensionTok{compiled}\NormalTok{ Rcpp attributes }
\ExtensionTok{edd@rob}\NormalTok{:~/git$}
\ExtensionTok{edd@rob}\NormalTok{:~/git$ mv RcppCorels/ rcppcorels  # prefer lowercase directories}
\ExtensionTok{edd@rob}\NormalTok{:~/git$ }
\end{Highlighting}
\end{Shaded}

\hypertarget{appendix-2-a-minimal-srcmakevars}{%
\subsection{Appendix 2: A Minimal
src/Makevars}\label{appendix-2-a-minimal-srcmakevars}}

\begin{Shaded}
\begin{Highlighting}[]

\ExtensionTok{CXX_STD}\NormalTok{ = CXX11}

\ExtensionTok{PKG_CFLAGS}\NormalTok{ = -I. -DGMP -DSKIP_MAIN}

\ExtensionTok{PKG_LIBS}\NormalTok{ = }\VariableTok{$(}\ExtensionTok{LAPACK_LIBS}\VariableTok{)} \VariableTok{$(}\ExtensionTok{BLAS_LIBS}\VariableTok{)} \VariableTok{$(}\ExtensionTok{FLIBS}\VariableTok{)}\NormalTok{ -lgmpxx -lgmp}
\end{Highlighting}
\end{Shaded}

\hypertarget{appendix-3-a-placeholder-wrapper}{%
\subsection{Appendix 3: A Placeholder
Wrapper}\label{appendix-3-a-placeholder-wrapper}}

\begin{Shaded}
\begin{Highlighting}[]

\PreprocessorTok{#include }\ImportTok{"queue.h"}

\PreprocessorTok{#include }\ImportTok{<Rcpp.h>}

\CommentTok{/*}
\CommentTok{ * Logs statistics about the execution of the algorithm and dumps it to a file.}
\CommentTok{ * To turn off, pass verbosity <= 1}
\CommentTok{ */}
\NormalTok{NullLogger* logger;}

\CommentTok{// [[Rcpp::export]]}
\DataTypeTok{bool}\NormalTok{ corels() \{}
  \ControlFlowTok{return} \KeywordTok{true}\NormalTok{;                  }\CommentTok{// more to fill in, naturally}
\NormalTok{\}}
\end{Highlighting}
\end{Shaded}


\end{document}